\documentclass{article}
\usepackage{spconf,amsmath,graphicx}
\usepackage{multirow,tabularx,booktabs,array}
\usepackage{color,soul}
\usepackage{xcolor}
\usepackage{subfigure}
\usepackage{amsmath,amssymb}
\usepackage{enumitem}
\usepackage{changepage}
\title{Clustering and Mining Accented speech for Inclusive and Fair Speech Recognition}
\name{Jaeyoung Kim, Han Lu, Soheil Khorram, Anshuman Tripathi, Qian Zhang, Hasim Sak}
\address{\{jaeykim, luha, soheilkhorram, anshumant,  zhaqian, hasim\}@google.com\\Google Inc., USA}

\begin{document}
\ninept
\maketitle
\begin{abstract}

% Automatic speech recognition (ASR) has shown great success 
% improved a lot with many breakthrough.
% But, ASR usually works best for canonical speech. 
% For example, ASR often significantly degrades with heavily accented speech. 

Modern automatic speech recognition (ASR) systems are typically trained on more than tens of thousands hours of speech data, which is one of the main factors for their great success. However, the distribution of such data is typically biased towards common accents or typical speech patterns. As a result, those systems often poorly perform on atypical accented speech.
In this paper, we present accent clustering and mining schemes for fair speech recognition systems which can perform equally well on under-represented accented speech. For accent recognition, we applied three schemes to overcome limited size of supervised accent data: supervised or unsupervised pre-training, distributionally robust optimization (DRO) and unsupervised clustering. Three schemes can significantly improve the accent recognition model especially for unbalanced and small accented speech. Fine-tuning ASR on the mined Indian accent speech using the proposed supervised or unsupervised clustering schemes showed 10.0\% and 5.3\% relative improvements compared to fine-tuning on the randomly sampled speech, respectively.

\end{abstract}
\vspace{-3pt}
\section{Introduction}
\label{sec:intro}
\vspace{-3pt}

Recent automatic speech recognition (ASR) systems have shown great success on diverse acoustic and linguistic conditions due to huge increase of model parameters and speech data. Typically, Modern ASR systems with several hundreds of thousands parameters can be easily trained on tens of thousands hours of speech data. The trained ASR systems will work as expected as long as target domains are well matched to training data.
However, the distribution of speech data is typically focused on standard canonical speech patterns. As a result, ASR systems trained on such data distribution often perform poorly on the unseen or atypical accented speech. 

The main challenge for recognizing accented speech is the lack of sufficient training data for these accents. It is very expensive and time-consuming to manually collect accented speech data especially when the target accent is rare and not often spoken. Therefore, well performing accent recognition systems would be important for systematically clustering and mining accented speech.
% Self-supervised pre-training method recently showed huge success in learning latent representations from unlabeled data and significantly improved ASR performance by fine-tuning small labeled speech data.

There has been extensive studies on joint learning of speech and accent recognition with multi-task method~\cite{rao2017multi,ghorbani2019leveraging,jain2018improved,li2018multi,zhang2021e2e,yang2018joint}.
For the most of the prior works, accent recognition is used as an auxiliary task to assist ASR system to be aware of accents in input audio speech. However there has been few research on investigating accent recognition models~\cite{berkling1998improving,jiao2016accent}. The most of works were evaluated as a part of challenges or use their well-defined accented datasets which are not usually available now.
Accented speech is rare and typically severely imbalanced between different accents. Furthermore, accent labels are not stable and sometimes severely corrupted. 

In this paper, we investigate an accent recognition model when we are given severely imbalanced accented datasets and accent labels are not reliable. We applied three schemes to address and resolve issues discussed above:

\noindent
\textbf{Pre-Training}: 
A pre-trained model can avoid learning the spurious features such as speaker identity by learning hidden speech representation from huge speech dataset. Both pre-training based on supervised and unsupervised training showed significant improvement on accent recognition.

\noindent
\textbf{Distributionally Robust Optimization (DRO)}: 
Group DRO minimizes the empirical risk of the worst-performing group, instead of minimizing the average empirical risk, which can avoid overfitting groups with smaller datasets. Group DRO effectively reduced accuracy variance between accents.

\noindent
\textbf{Unsupervised Clustering}: 
Unsupervised clustering is used to recognize unseen accents during the model training stage. K-means algorithm is applied to only update centroids' locations by fixing the weights for the trained model.

An ASR model based on Transformer-Transducer is fine-tuned on the mined Indian accents from the proposed supervised and unsupervised accent models. They showed 10.0\%  and 5.3\% relative WER improvements on the Indian accent compared to randomly sampled accents, respectively.

% \begin{itemize}
% \item Pre-Training:
% \end{itemize}

% \textbf{Pre-Training:} 

% on the data imbalance between different accents, impurity of accent labels and its impact on accent recognition, and mining new accented speech to fine-tune accented ASR. 

% Accented speech data is rare and typically severely imbalanced between different accents. There are many  

% especially for mining new accented speech to finetune and improve accented ASR. 

% Compared to joint learning, there has been smaller studies for accent recognition system~\cite{berkling1998improving,jiao2016accent}. Especially, accent recognition 

% In this paper, we present  accent recognition system

% \section{Related Work}
% \label{sec:related work}

\section{Accent Recognition}
\label{sec:cross training}

\begin{figure}[h]
\centering
\includegraphics[width=0.8\columnwidth]{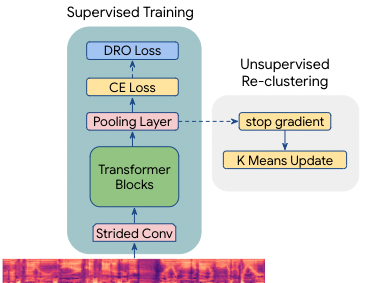}
\caption{Overview of Training Accent Recognition Models}
\label{fig:accent_model}
\end{figure}

\begin{figure*}[h]
\centering
\includegraphics[width=2.0\columnwidth]{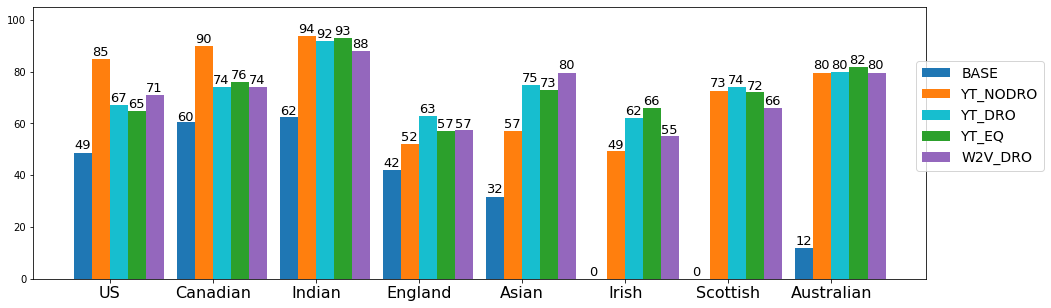}
\caption{Comparison of Accent Recognition Models}
\label{fig:prediction_result}
\end{figure*}

Figure~\ref{fig:accent_model} shows the proposed accent recognition model. The model consists of convolutional layers and Transformer block. The convolutional layers sub-sample log-mel spectrograms with strides of 4. The strided convolution output is fed to the Transformer block and its output is pooled across time for the accent representation. There are several pooling methods. Among them, average and maximum pooling were evaluated, which showed no significant difference in accent prediction. In this paper, average-pooling was used.  

Figure~\ref{fig:accent_model} presents two stages of model training: supervised cross entropy (CE) with distributionally robust optimization (DRO) and unsupervised re-clustering by online K-means algorithm. DRO can effectively deal with data imbalance between different accents, which will be explained at section~\ref{subsec:dro} in detail. The unsupervised re-clustering is useful when supervised accent labels are noisy. Online K means algorithm is used for computing new centroids for accent re-clustering.

\subsection{Pre-Training}

One of the challenges for training an accent recognition model is to prevent the model from correlating features other than speech accents to the predictions. For example, the model can utilize gender, speaker tone, channel characteristics or background noise to distinguish accents especially when accent data is not large enough. Table~\ref{tab:mozilla} summarizes the number of utterances between 8 English accents provided by Mozilla Common Voice data. Among all the accents, Asian or Irish accents are much smaller than US accent and hence a recognition model can be susceptible to correlate based on spurious features such as speaker identity.

A pre-trained model can avoid learning the biases by learning hidden speech representation on huge supervised or unsupervised speech data. For supervised pre-training, we utilize RNN-T ASR model trained on 400K supervised Youtube data. The audio encoder block in Youtube RNN-T model is used for initializing an accent recognition model. The audio encoder block has the same structure as the accent recognition model in Figure~\ref{fig:accent_model}:2 convolutional layers and 20 Transformer layers. For a self-supervised model, wav2vec training is applied on one million unsupervised Youtube speech data. One difference on wav2vec model is that the softmax output for accent recognition is derived from $10^{\text{th}}$ layer of wav2vec model. Typically, self-supervised models trained on input audio feature have better feature abstraction in the middle of layers because layers closer to output should reconstruct raw audio features. Two pre-trained models are evaluated at section~\ref{sec:result_accent}. Both of them showed huge improvements over a randomly initialized model.

\subsection{Distributionally Robust Optimization (DRO)}

Group DRO \cite{sagawa2019distributionally} was first introduced on computer vision and natural language processing applications that tries to make models robust to different group shifts at test time. It achieves that by minimizing the empirical risk of the worst-performing group, instead of minimizing the average empirical risk (i.e. ERM) at training time. Such optimization avoids learning correlations between input features and output targets that hold on average, but do not generalize to broader use cases. In practice, to train DRO efficiently, \cite{sagawa2019distributionally} proposed an online algorithm that tracks the loss for each training group to form a distribution over different groups. Such distribution is used to scale the losses from different groups to optimize the model.

Publicly available accented speech data is usually imbalanced between different accent groups due to the difficulty of collecting rare accented speech. Table~\ref{tab:mozilla} showed US accent is the most common and constitutes almost the half of total utterances. Even the second largest accent group (i.e. England) is only around one third of US accent. A neural network model trained to minimize the average empirical risk on a highly imbalanced training dataset tends to work better on larger accent groups (e.g. US) but often perform poorly on the smaller rare accent groups. A widely used technique to alleviate such performance gap due to class imbalance is to sample equally from each training class to have higher exposure to rare classes. Although this is effective in some cases, this tends to overfit easily and does not generalize well for most of the accent groups. On the contrary, DRO scales the loss of worst-performing accent group more in addition to sampling each accent group equally. This prevents the model from further optimizing on smaller groups even when the corresponding losses are already small. DRO showed the smallest prediction variance between different accents without hurting mean prediction accuracy at Table~\ref{tab:yt}. 

\label{subsec:dro}
\subsection{Unsupervised Clustering}

\begin{figure*}
    \centering
    \subfigure[]{\includegraphics[width=0.33\textwidth]{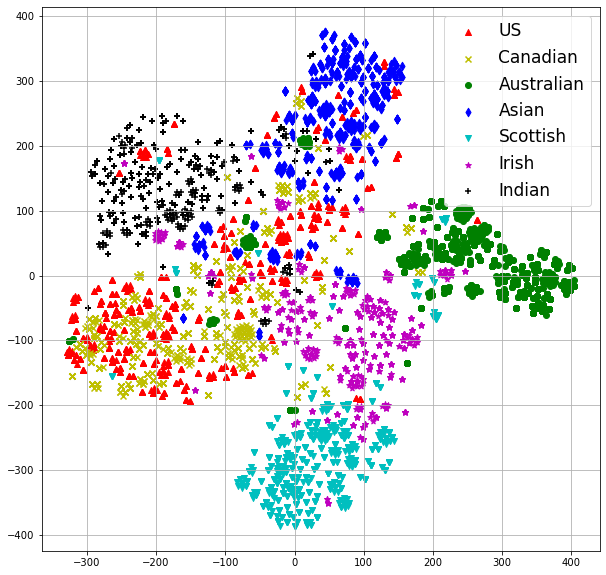}} 
    \subfigure[]{\includegraphics[width=0.33\textwidth]{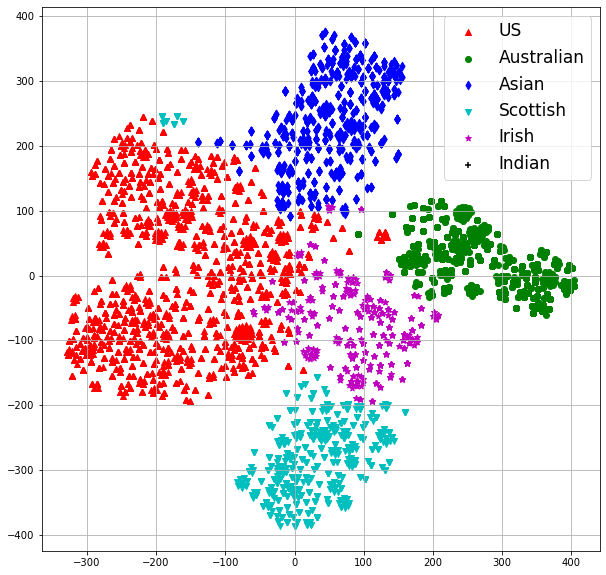}} 
    \subfigure[]{\includegraphics[width=0.33\textwidth]{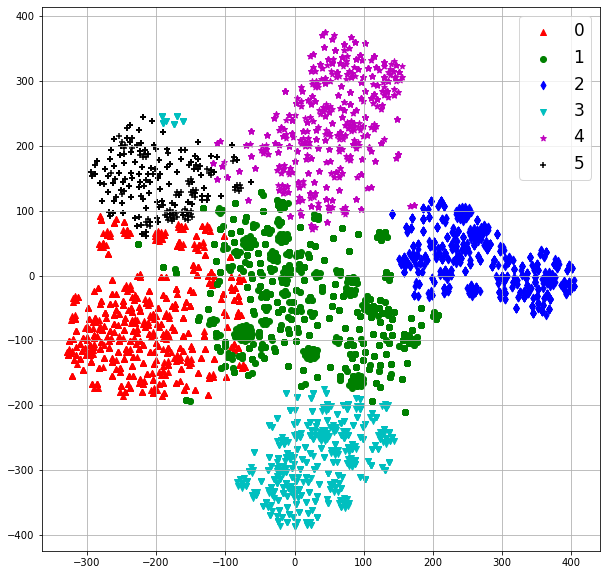}}
    \caption{TSNE Projection of the Accent Recognition Model: (a) Ground Truth (b) Model Prediction (c) K Means Clustering}
    \label{fig:tsne_embedding}
\end{figure*}

Figure~\ref{fig:tsne_embedding} shows TSNE plots for different testing accent embeddings extracted from the pooling layer output of the trained accent recognition model. Each plot at Figure~\ref{fig:tsne_embedding} depicts different clustering on the same set accent embeddings. For example, Figure~\ref{fig:tsne_embedding} (a) shows accent groups by ground-truth labels and Figure~\ref{fig:tsne_embedding} (b) presents clustering by the model's softmax predictions. There are a couple of things to note for the accent recognition model. First, since Canadian and US accents are close together, they are treated as the same accent during model training. Second, the accent recognition model was trained without Indian accents in order to test the generalization capability for the unseen accent clustering.

There are several observations on the ground-truth TSNE plots. First, TSNE plot shows nicely separated embedding groups. Especially, the Indian accents are clustered well from other accents, which suggests the trained accent model generalized well to unseen accents.
Second, some accents such as US, Asian and Irish are mixed on the center region without clear boundaries. It is because the accent labels are collected based on speakers' location and therefore sometimes they do not match with the speakers' true accents: noisy accent labels. For example, we sampled US accent utterances which overlap with Indian accent embedding location in Figure~\ref{fig:tsne_embedding} (a). We listened to each sampled utterance and confirmed all of them are closer to Indian accents than US ones.

Figure~\ref{fig:tsne_embedding} (b) shows accent groups predicted by the recognition model. The model clearly predicted well for accents seen during training but it did not correctly recognize Indian accents. The model cannot predict Indian accent because its softmax output only contains accents seen from training data. However, as explained before, the model showed good Indian accent clustering in the TNSE plot.

We propose the unsupervised accent clustering based on K-means algorithm. The main idea is that K-means algorithm only updates centroids' location and does not update the trained accent recognition model. The number of K-means centroids can be set flexibly large enough to cover unseen accents. Figure~\ref{fig:tsne_embedding} (c) showed K-means clustering when the number of centroids are 6. The Indian accents are correctly recognized compared to the ground-truth plot.

% Accent embeddings were less affected by noisy labels due to pre-training initialization from huge speech data. However, the model prediction especially for the softmax layer is significantly affected: US or England accent predictions are mixed on the center and upper right embedding groups at Figure~\ref{fig:tsne_embedding} (b). 

% Due to the observations mentioned above, we applied online K-means clustering algorithm to re-cluster learned accent embeddings. The goal of the re-clustering is to find out the centroids of each accent, which can later be used to re-cluster the training data. The re-clustered accent labels can be used to re-train the accent recognition model by filtering out noisy labels from training set and boosting models with cleaner ones. The details about cleaning, re-training and the precision/recall trade-offs for the retrained model would be described in section~\ref{sec:k-means}.

% The K-means algorithm only updates centroids' locations and does not update the trained accent recognition model. Otherwise, the model can easily collapsed to the trivial states. The unsupervised clustering can be robust to label mistakes because they are not trained by accent labels. 

% The another benefit for the re-clustering is its robustness to unseen accents. A softmax layer can only predict predefined accent labels. On the contrary, the number of K-means centroids can be set flexibly large enough to cover future unseen accents. It is not necessarily one-to-one matched to the size of training accent labels.

Here we describe the details of the online K-means algorithm we used. We followed K-means++ algorithm for centroid initialization:
\begin{enumerate}
    \item Randomly select data point $x \in X$ as the first centroid $c_1$.
    \item Each new centroid $c_i$ is sampled based on the distance \\ probability:$\frac{D(x)^2}{\sum_{x \in X} D(x)^2}$ where $D(x)$ is the distance \\
    to the closest centroid from data $x$.
    \item Repeat 2 until initializing all of centroids.
\end{enumerate}
Centroids are updated as follows:
\begin{itemize}
    \item Compute new centroids: $\hat{c}_i = \frac{1}{|C_i|}\sum_{x \in C_i} x$
    \item EMA updates: $c^t_i = \alpha c^{t-1}_i + (1-\alpha) \hat{c}_i$
\end{itemize}
where $C_i$ is the set of data points belonging to the $c_i$ centroid, $|C_i|$ is the cardinality of the set, EMA is exponentially moving average and $\alpha$ is an EMA update weight.

% \begin{figure*}[t]
% \centering
% \begin{subfigure}
%     \centering
%     \includegraphics[width=.95\linewidth]{figs/tsne_gt_gdro.png}  
%     \caption{}
%     \label{SUBFIGURE LABEL 1}
% \end{subfigure}
% \begin{subfigure}
%     \centering
%     \includegraphics[width=.95\linewidth]{figs/tsne_predict.png}  
%     \caption{}
%     \label{SUBFIGURE LABEL 2}
% \end{subfigure}
% \begin{subfigure}
%     \centering
%     \includegraphics[width=.95\linewidth]{figs/tsne_kmeans.png}  
%     \caption{}
%     \label{SUBFIGURE LABEL 3}
% \end{subfigure}
% \caption{FIGURE CAPTION}
% \label{FIGURE LABEL}
% \end{figure*}

%\subsubsection{Self-Training}
%\subsection{K-Means Label Smoothing}
\vspace{-3pt}

\section{Experiment and Results}
\label{sec:experiments}
\vspace{-3pt}

\subsection{data}

We used three separate datasets: Mozilla Common Voice~\cite{mozilla}, Youtube supervised (YT-L) and Youtube unsupervised (YT-U) datasets and LibriSpeech. The Mozilla Common Voice is a speech dataset collected from voluntary user submissions based on various public domain sources. The dataset contains various English accents based on different countries. The total hours for training data with accent labels are around 700 hours and the total number of English accents are 14. However, due to dataset imbalance between accents, some accents only have a few hours of training data. In this paper, we merged accent groups to have a total of 8 accents as in Table~\ref{tab:mozilla}. Youtube is a set of large audio datasets of anonymized Youtube Videos collected in accordance with Google Privacy and AI principles~\cite{googleprivacy, google_ai_principles}. Youtube supervised (YT-L) is roughly 350k hours of segmented, weekly-labeled audio, combined with 1000 hours of manually labeled data. Youtube unsupervised (YT-U) is a 900k hours of unlabeled audio dataset. It is first randomly collected from 3M hours Youtube videos of interviews and lectures, and then filtering non-speech parts later. Finally, LibriSpeech~\cite{panayotov2015librispeech} corpus is a read speech data based on audio books and consists of 1000 hours of training and test sets.

\begin{table}[t]
  \centering
  \begin{tabularx}{0.45\textwidth}{l || r r | r r}
    \toprule
     &\multicolumn{2}{c|}{Train} & \multicolumn{2}{c}{Test} \\
    Accents & Utts      & Speakers & Utts   & Speakers \\
    \hline \hline \\
    US      & 220000    &	4135    & 1590  &   750  \\
    Indian  & 73000	&1026	&708	&347  \\
    England & 75000	&1327	&401	&202    \\
    Canadian & 39000&	532&	152	&68    \\
    Australian & 31000	&364&	77	&36    \\
    Asian & 9820&	214	&116&	53    \\
    Irish & 5867 &	88 &	150 &	33 \\
    Scottish & 9864 &	81 &	119	 & 23 \\
    % \hspace{-2.5pt}\footnotesize{\textbf{Baseline Models}} & & & & & \\
    % random init. & .45 & \per{1.9} & \per{2.0} & \per{4.2} & \per{4.5}  \\
    \bottomrule
  \end{tabularx}
  \caption{Mozilla Common Voice English Accent Dataset.
}
  \label{tab:mozilla}
  \vspace{-12pt}
\end{table}

\begin{table}[t]
  \centering
%   \setlength{\tabcolsep}{7pt}
%   \begin{tabularx}{0.45\textwidth}{l || c | c | c }
  \begin{tabularx}{0.45\textwidth}{l || c | c | c }
    \toprule
    Method & INIT & ACC MEAN & ACC STDEV \\
     \hline \hline \rule{0pt}{12pt}
    BASE & Random & 32.2\% & 25.6  \\
    YT & Youtube & 72.4\% & 17.6  \\
    YT\_DRO & Youtube & \textbf{73.4}\% & \textbf{9.8}  \\
    YT\_EQ & Youtube & 73.0\% & 11.1  \\
    W2V\_DRO & W2V2 & 71.4\% & 11.4  \\
    \bottomrule
  \end{tabularx}
  \caption{Accent prediction accuracy (\%) comparison between models with different initialization and optimization.}
  \label{tab:yt}
  \vspace{-12pt}
\end{table}

\subsection{Result on Accent Recognition}
\label{sec:result_accent}

The accent recognition model has 20 Transformer layers based on bidirectional transformer-XL~\cite{dai2019transformer} and 2 2d-convolutional layers with total of stride 4. There are total of 5 different settings: Base, $\text{YT}\_\text{NODRO}$, $\text{YT}\_\text{DRO}$, $\text{YT}\_\text{EQ}$ and $\text{W2V}\_\text{DRO}$. Base model is trained on Mozilla Common voice dataset with random initialization. $\text{YT}\_\text{NODRO}$ is initialized from a RNN-T ASR model trained on YT-L.  $\text{YT}\_\text{DRO}$ applied DRO optimization to $\text{YT}\_\text{NODRO}$, $\text{YT}\_\text{EQ}$ is to sample unbalanced accents equally, and $\text{W2V}\_\text{DRO}$ has an wav2vec2.0 initialization~\cite{baevski2020wav2vec} trained on YT-U dataset.

Figure~\ref{fig:prediction_result} compares accent prediction accuracies between different settings. For the BASE model, most of accent predictions show poor performance. Especially, Irish or Scottish accent predictions showed $0\%$ accuracies due to the lack of accented data. The $\text{YT}\_\text{NODRO}$ initialized from Youtube RNN-T model showed significant improvements on almost every accent predictions. However, the issue on the model is that prediction performance is significantly different between accents. For example, US, Canadian and Indian accents are around $90\%$ accuracies but Asian, Australian and England are less than $50\%$. Furthermore, Mozilla accent labels are weekly-annotated using indirect meta information such as residence location. Too high accuracies on the test set especially for US or Canadian accents can be the indication of overfitting to speaker information, channel condition or background noise other than accent information. Distributionally robust optimization (DRO) reduced accuracy variance between accents for $\text{YT}\_\text{DRO}$. There are degradation for high performing accents such as US and Canadian but the rest of accent accuracies uniformly improved. Similarly, DRO model based on wav2vec initialization showed reduced variance between accents. Table~\ref{tab:yt} shows accuracy mean and variance over all accent groups. Mean accuracies are similarly good for both DRO and the equal sampling scheme ($\text{YT}\_\text{EQ}$). However, DRO showed smaller standard deviation than EQ. 

\subsection{Fine-Tuning ASR on the Mined Accented Speech}
% compare model between base vs self-training vs k-means label smoothing
\begin{table}[t]
  \centering
%   \setlength{\tabcolsep}{7pt}
%   \begin{tabularx}{0.45\textwidth}{l || c | c | c }
  \begin{tabularx}{0.49\textwidth}{l l|| c | c  }
    \toprule
    Accent &ASR Model &  WER & WERR (\%)  \\
     \hline  \rule{0pt}{12pt}
    &BASE & 36.8 & 0.0   \\
    &No Indian YT-L & 16.2 & 56.0   \\
    Indian &Random YT-L& 15.2 & 58.7  \\
    &Indian YT-L& 13.7 & 62.7  \\
    &K-Means Indian YT-L & 14.4 & 60.9  \\
    \hline
    \rule{0pt}{12pt}
    Accent &ASR Model &  WER & WERR (\%)  \\
    \hline  \rule{0pt}{12pt}
    &BASE & 24.2 & 0.0   \\
    & No Indian YT-L & 11.93 & 50.7   \\
    US & Random YT-L& 11.88 & 50.9  \\
    & Indian YT-L& 11.87 & 50.9  \\
    & K-Means Indian YT-L & 11.88 & 50.9  \\
    \bottomrule
  \end{tabularx}
  \caption{Fine-tuning ASR on the mined accented speech}
  \label{tab:ft}
  \vspace{-12pt}
\end{table}
In this section, an ASR model based on Transformer-Transducer (T-T)~\cite{zhang2020transformer,yeh2019transformertransducer,tripathi2020transformer} is fine-tuned on the accent speech mined by the proposed accent recognition model. The audio encoder consists of two strided convolutions followed by 20 full-attention Transformer layers. The label encoder has a couple of streaming Transformer layers. 80-dimentional log-mel filterbanks are acoustic features to the audio encoder and SpecAugment and variational noise are added to the model input and parameters for robust training. 

The BASE model is initially trained on LibriSpeech, which is then fine-tuned on the different mined accented speech as in Table~\ref{tab:ft}. For Random YT-L, fixed size of YT-L speech data is randomly sampled. For No Indian YT-L, Indian accents were removed from random samples from YT-L. For Indian YT-L, Indian accents are mined by the accent model training with Indian accents. Lastly, for K-Means Indian YT-L, Indian accents are mined by the model trained without Indian accents. For a fair fine-tuning environment, all the fine-tuning setups have similar number of utterances. Since fine-tuning only on the specific accents frequently degrades other accents, US accented data from Mozilla Common Voice was mixed with mined speech data.

% Accent mining happens on YT-L dataset. For a fair fine-tuning environment, all mining setups have similar size of mined accents.

In Table~\ref{tab:ft} fine-tuning on the mined Indian accents (Indian YT-L) showed significant improvements on the Indian accent recognition: 10.0\% and 15.4\% relative improvement over the random YT-L and random YT-L without the Indian accents. The unsupervised Indian accent clustering (K-Means Indian YT-L) also improved the Indian accent recognition: 5.3\% and 11.1\% relative improvements over Random YT-L and No Indian YT-L although the gain decreased compared to the supervised clustering. The WER gap between them would be because the Indian accent recognition for the unsupervised clustering is lower than supervised one, therefore, mined dataset would contain less Indian utterances. 

Evaluation on the US accents for all the fine-tuned ASR models showed similar WER. We did not see any degradation from fine-tuning on the mined Indian accent when small US accents were trained together.

% For the England accent, WER improvement decreased compared to the Indian accent: 4.1\% and 5.4\% for random selection with and without England accents. The reason for lower improvement for the England accent could be the lower precision of mining English accents compared to mining Indian ones.
\vspace{-5pt}
\section{Conclusion}
\label{sec:conclusion}

In this paper, we showed the new framework for improving accent recognition models. Pre-training and group DRO schemes were applied to overcome unbalanced and small accented speech. We also evaluated supervised and unsupervised clustering of accented speech and their application to fine-tuning ASR models. We have shown that the proposed accent recognition model can generalize well on  the unseen accented speech by fine-tuning ASR on the mined accented speech. As a future work, we will investigate how to reduce the gap between supervised and unsupervised clustering to improve its generalization on the unseen accent recognition.  

% \vspace{-5pt}

% Explain what will happen if we do not use any of the components.

\bibliographystyle{IEEEbib}
\bibliography{refs}

\end{document}